# Coupled Electron-Nuclear Dynamics Induced and Monitored with Femtosecond Soft X-ray Pulses in the Amino Acid Glycine

*David Schwickert[1], Andreas Przystawik[1], Dian Diaman[1], Detlef Kip[2], Jon P. Marangos[3], and Tim Laarmann[1,4]\**

[1]Deutsches Elektronen-Synchrotron DESY, Notkestr. 85, 22607 Hamburg, Germany.

[2]Faculty of Electrical Engineering, Helmut Schmidt University, Holstenhofweg 85, 22043 Hamburg, Germany.

[3]Department of Physics, Imperial College London, Prince Consort Road, London SW7 2AZ, United Kingdom.

[4]The Hamburg Centre for Ultrafast Imaging CUI, Luruper Chaussee 149, 22761 Hamburg, Germany.

ABSTRACT: The coupling of electronic and nuclear motion in polyatomic molecules is at the heart of attochemistry. The molecular properties, transient structures and reaction mechanism of these many-body quantum objects are defined on the level of electrons and ions by molecular wave functions and their coherent superposition, respectively. In the present contribution we monitor nonadiabatic quantum wave packet dynamics during molecular charge motion by reconstructing both, the oscillatory charge density distribution and the characteristic time-dependent nuclear configuration coordinate from time-resolved Auger electron spectroscopic data recorded in previous studies on glycine molecules [*Sci. Adv.* **2022**, *8*, eabn6848]. The electronic and nuclear motion on the femtosecond timescale was induced and probed in kinematically complete soft x-ray experiments at the FLASH free-electron laser facility. The detailed analysis of amplitude, instantaneous phase and instantaneous frequency of the propagating many-body wave packet during its lifecycle provides unprecedented insight into dynamical processes beyond the Born-Oppenheimer approximation. We are confident that the refined experimental data evaluation helps to develop new theoretical tools to describe time-dependent molecular wave functions in complicated, but ubiquitous, non-Born-Oppenheimer photochemical conditions.

INTRODUCTION

The Born-Oppenheimer approximation is a cornerstone for our understanding of molecular structure and dynamics in a quantum mechanical picture[1, 2]. It allows to simplify the time-dependent total molecular wave function:

$$\Phi(\boldsymbol{r}, \boldsymbol{R}, t) = \Phi(\boldsymbol{r_1}, \boldsymbol{r_2}, \dots, \boldsymbol{r_N}, \boldsymbol{R_1}, \boldsymbol{R_2}, \boldsymbol{R_M}, t), \qquad (1)$$

where $\boldsymbol{r}, \boldsymbol{R}$ are vectors pointing towards the electronic and nuclear coordinates of *N* electrons and *M* nuclei comprising a polyatomic molecule. This mathematical object describes the properties of the molecule and its response to external stimuli perturbing the many-body quantum system. The thus induced quantum dynamical evolution is dictated by the time-dependent Schrödinger equation (TDSE):

$$i\hbar\frac{\partial}{\partial t}\Phi(\bm{r},\bm{R},t) = \hat{H}(\bm{r},\bm{R})\Phi(\bm{r},\bm{R},t), \qquad (2)$$

and driven by the non-relativistic Hamiltonian $\hat{H}(\bm{r},\bm{R})$ including all the interactions at work. We note in passing that the non-relativistic description is a valid approximation for many processes relevant in organic chemistry since (i) the structural changes take place in molecular compounds built of low-Z elements (H, C, O, N, …) and (ii) their interaction with soft X-rays as in the present study generates photoelectrons of rather low kinetic energy below a few keV.

The difficulties to solve the TDSE scale with the number of particles comprising the molecule, the degree of electronic correlation that has to be considered, and last but not least with the characteristic timescale for electronic and nuclear motion related to the energy separation between the corresponding quantum mechanical eigenstates[3]. Within the Born-Oppenheimer approximation and in the absence of perturbations the large but finite quantum system is basically described by a single nuclear wave function, whose time-dependence is determined by a single electronic eigenstate neglecting the possibility that the nuclear motion can cause a change in the electronic state[4]. In other words, the nuclei evolve adiabatically on a potential energy surface, which depends on the 3*M*-6 nuclear degrees of freedom. The multidimensional potential energy landscape is derived by calculating the electronic eigenstate energies with the nuclear configuration coordinate $\bm{R}$ entering as a parameter. In this scenario, there is no possibility for the nuclei to give energy to the electrons for population transfer from one electronic eigenstate to another even if the state is located close by in the electronic energy level diagram.

It is obvious, why the Born-Oppenheimer approximation is incredibly successful when describing chemical processes taking place in the electronic ground state. The typical energy separation of electronic eigenstates in molecules is on the order of a few eV with 1 eV corresponding to a thermal energy for temperature ~11600 K. Therefore, at room temperature it is extremely unlikely that nuclear motion is energetic enough to trigger an electronic transition[2]. However, the situation changes dramatically upon fast inner-valence photoionization using femtosecond pulses, where the Born-Oppenheimer approximation breaks down. This is the situation in the present study. In particular two aspects need to be considered here:

(i) As the energy of the ionic states increases the effect of electron correlation becomes more important. This typically leads to the formation of shake-up satellites and ultimately for deeply bound states to the complete breakdown of the molecular orbital picture[5]. At least two-hole–one-particle (2h1p) configurations need to be included in the theoretical simulations of the cationic electronic structure, because the sudden removal of an electron from a particular orbital is accompanied by the excitation of another electron to an initially unoccupied orbital.

(ii) The impulsive ionization results in the formation of a coherent superposition of electronic states in the cation[6]. Ionization paths leading to the same photoelectron energies, but leaving behind different cationic states, interfere and trigger coherent quantum motion of the remaining correlated $N–1$ electrons. The timescale of the resulting oscillatory variation of related observables, such as charge density, depends on the energy separation of the coherently coupled molecular states[7].

Thus, a key control parameter of the ionizing short-wavelength laser pulse is the coherent spectral bandwidth, which affects the dynamic evolution of the molecular wave function and its superposition, respectively. In particular, it allows to prepare the many-body quantum system in a well-defined coherent superposition of specific electronic states that matches characteristic frequencies of nuclear dynamics. This was exactly the experimental strategy applied in a recent study enabling a detailed analysis of nonadiabatic electron-nuclear dynamics, which follows the prompt ionization of the prototypical biomolecule glycine with ultrashort femtosecond (fs) pump pulses in the soft x-ray spectral range[8]. The overarching goal to watch the birth, propagation, and fate of the resulting many-body quantum wave packet as a function of electronic and nuclear coordinates has been achieved by reconstructing the amplitude, the instantaneous phase and the instantaneous frequency of the modulus squared of the time-dependent total molecular wave function or more precisely of the excited coherent superposition of cationic eigenstates from time-resolved pump-probe data. These quantities are derived from further analysis of the experimental data published in ref.[8], which has been recorded at the free-electron laser (FEL) facility in Hamburg FLASH[9]. The corresponding findings discussed below are presented here for the first time.

EXPERIMENTAL METHODS

In order to track the back-and-forth propagation of the inner-valence electron hole across the molecular-ion backbone of glycine created by means of photoionization and the evolving coupled electron-nuclear dynamics, the soft x-ray FEL probe photon energy was set to a local, element-specific, site- and orbital selective core – inner valence shell transition below the carbon K-edge absorption. The experimental details of the single-color pump-probe study on glycine molecules in the gas-phase can be found in ref.[8]. In brief, the effusive molecular glycine beam source based on a resistively heated oven was operated at around 160°C, which results in a gas pressure on the order of $10^{-2}$ mbar in front of the nozzle. We estimated the target density in the FEL interaction zone to about 900 molecules/mm$^3$. The detection of the generated electrons and ions is realized using a magnetic-bottle electron spectrometer (MBES) and a time-of-flight (TOF) ion spectrometer with the detection axis being oriented perpendicular to the FEL and to the molecular beam direction. The positions of the detector and electrodes of the MBES are fixed, whereas the positions of the interaction zone (FEL focus and capillary position), ion TOF spectrometer and the permanent magnet of the MBES can be adjusted.

The selected electronic core – inner valence shell transition at a photon energy of 272.7 eV into the spatially extended inner-valence 10a′ molecular orbital of the glycine cation induces Auger electron emission that is detected in coincidence with the generated parent ions and the corresponding photoelectron with a binding energy of ~20 eV. Thereby, the time-delayed fs x-ray pulses probe the transient local charge (hole) density that has been created initially by a pump pulse of the same color at one specific carbon atom ($C_\alpha$). This is because the local core – inner valence shell absorption cross section, i.e. the detected Auger electron yield, is proportional to the time-dependent positive charge density. The basic experimental scheme flanked by ab-initio simulations has already been proposed by Cooper et al.[10] in 2014.

The key parameter controlling the initial ionization-induced dynamics is the rather small spectral bandwidth of the applied fs FEL pump pulse. The pulse spectrum has been characterized ($\Gamma = 0.37\%$) at a central wavelength of $\lambda = 4.55$ nm corresponding to the FEL photon energy of 272.7 eV. The selected spectral properties allow for resonant excitation of specific core – inner valence shell transitions, namely from the 5a′ orbital localized in the vicinity of the $C_\alpha$ atom into the 10a′ band energy region. It is important to note that the initial prompt photoionization with the radiation pulses of ~1 eV spectral bandwidth results in a coherent superposition of a series of cationic eigenstates with contributions from the inner-valence hole (1h) state and excited 2h1p configurations[8]. The induced coherence is monitored with fs resolution defined by the corresponding Fourier-limited Gaussian pulse duration of 1.8 fs (FWHM). However, we assume the FEL pulse to be slightly down-chirped resulting in a pulse duration of 3-5 fs (FWHM), which gives the order of magnitude of the available temporal resolution in the experiments. The general experimental scheme is sketched in Fig. 1.

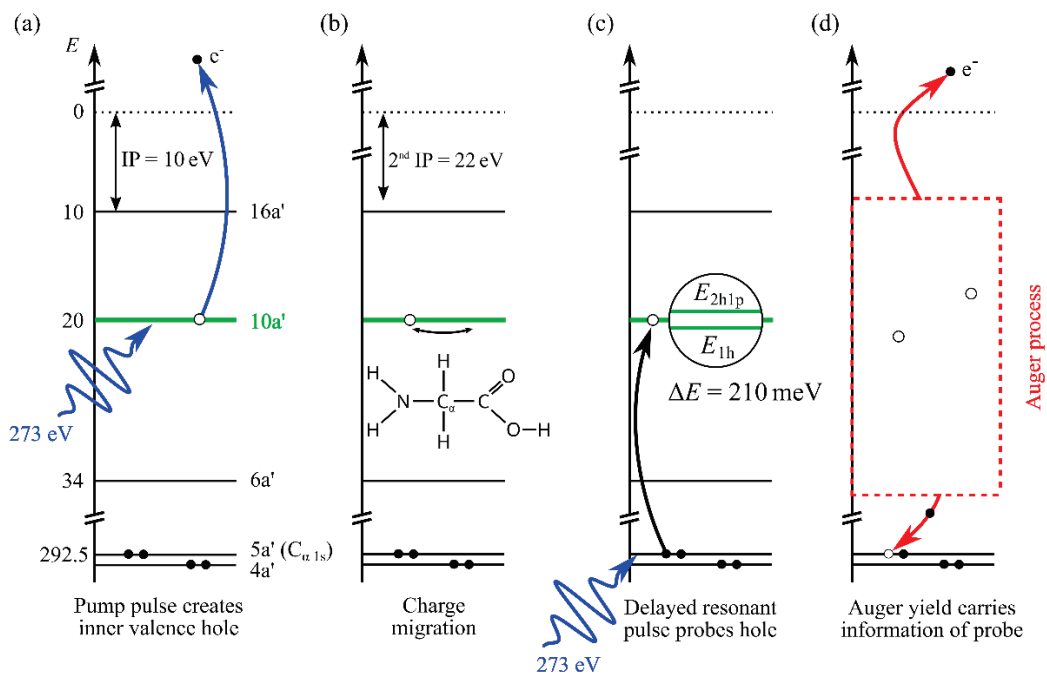

**Figure 1.** Time-resolved x-ray Auger electron spectroscopy of glycine. A single-color pump-probe scheme is applied to monitor nonadiabatic quantum wave packet dynamics initiated by photoelectron emission from the 10a′ molecular orbital of glycine molecules. The total cascade involves several different processes: (a) photoionization, (b) charge dynamics triggered by the excitation of a coherent superposition of the inner-valence hole state (1h) and a series of two-hole one-particle (2h1p) configurations populated during photoionization, (c) resonant carbon 1s core – inner valence shell transition and (d) Auger decay. The individual steps are depicted as a sequence of energy diagrams adopted from ref.[8].

## RESULTS AND DISCUSSION

The selection of ionization events according to the kinetic energy of emitted photoelectrons allows to address specifically cationic eigenstates in the 10a′ band energy region i.e., the states resulting from photoelectron emission from the 10a′ orbital with a binding energy of ~20 eV. The overall level splitting is on the order of 210 meV and comprising the inner-valence hole (1h) state and a series of excited 2h1p configurations mediated by electronic correlation. The probe-induced Auger electron cascade from C 1s core vacancies cover the whole extent of the studied kinetic energy range as shown in the recorded electron-electron coincidence map of detected electrons integrated over 175 fs pump-probe delay in Fig. 2.

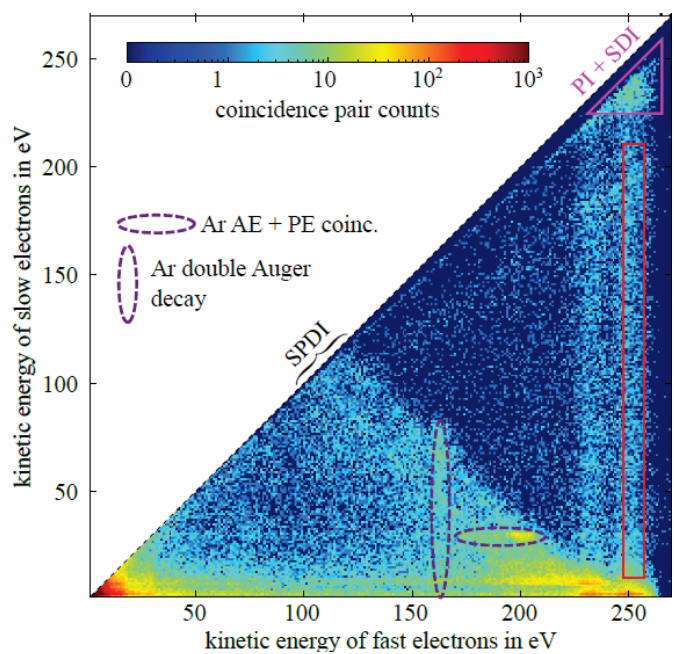

**Figure 2.** Two-electron coincidence spectroscopy of glycine with 273.0 eV photons. Photoelectron–Auger electron coincidences following the resonant soft x-ray transition (1s → 1h) are indicated (red rectangle). Valence electron emission from x-ray probe-induced

sequential double ionization (SDI) contributes to an uncorrelated background with respect to the initial x-ray pump-induced photoionization (PI) events. The map also shows single-photon double photoionization (SPDI) of glycine, as well as photoelectrons (PE) and Auger electrons (AE) originating from residual argon gas used for spectrometer calibration.

If two valence photoionization (PI) events happen sequentially on the same glycine molecule, the final two-hole (2h) dicationic state with two unbound continuum electrons is similar to that from Auger decay. In the special case, when the kinetic energy of such electrons happens to overlap with that of Auger electrons as marked in the upper right triangle $E_{kin,1} \gtrsim E_{kin,2} \gtrsim 225$ eV, the two processes become virtually indistinguishable by their end products. The photoelectron spectrum of the second photoelectron is generally shifted by the difference between the first and second ionization potentials but may also cover the extends of the single ionization spectrum due to ionization of high-lying Rydberg states of the cation as discussed in ref.[8]. In addition to the former sequential double photoionization (SDI) process, single-photon double photoionization (SPDI) of glycine is also observed resulting in diagonal structures perpendicular to the auto-coincidence diagonal. In a quasi-classical picture, a photon can only be absorbed by one electron but simultaneous ejection of a second photoelectron may occur if energy is transferred through the correlated motion of electrons. Their kinetic energies can be at most 241 eV, due to the molecules' double-ionization potential of 32 eV and the FEL photon energy of 273 eV. The cross-section of this high-order process can be up to four orders of magnitude lower than that of direct photoionization[11, 12], which depends on the photon energy and the involved electron binding energies. The amount of energy shared and thus the splitting ratio of the electrons' kinetic energies is a complex quantum many-body problem and relates to the question whether the secondary electron is ejected through a shake-up or knockout mechanism, which is still subject of active research[11], but not in the focus of the present study. We note in passing that the coincidence peak of Auger electrons between 170 eV and 207 eV and corresponding photoelectrons at 29 eV (horizontal ellipse in Fig. 2) originates from residual argon gas, which was used for spectrometer calibration prior to the glycine measurements. Likewise, the coincidence events at around 163 eV and < 80 eV (vertical ellipse) are a result of double Auger decay in argon[13]. These ionization events do not affect the following analysis, since they are filtered out based on their mass-to-charge ratio of 20 u/$e$ for $Ar^{2+}$ ions.

The most relevant ionic interaction product when studying the many-body quantum wave packet as a function of electronic and nuclear coordinates in glycine molecules is the intact mother ion. In the following, only events were selected (red box in Fig. 2), which included at least one electron from the 10a′ orbital (247 eV to 257 eV at $E_{ph}$ = 272.7 eV), at least one Auger electron without SDI contributions (10 eV < $E_{kin}$ < 210 eV) and at least one $Gly^{2+}$ ion ($m/q$ = (37.5 ± 4) u/$e$). The above kinetic energy selection also excludes electrons resulting from excitation of a C 1s state to a Rydberg state by the pump pulse followed by Auger decay of the core-excited neutral Gly molecule even without interaction with the probe pulse, as well as those events induced by subsequent valence or Rydberg photoionization by the probe pulse. In the analysis, covariance counting (photoelectron-Auger electron-photoion covariance) was favored over coincidences, because of the occurrence of multiple events per shot resulting in too many false coincidences. In order to primarily evaluate ionization events induced by single-mode, few-fs self-amplified spontaneous emission (SASE) pulses, only those with pulse energies below 2 µJ have been selected. Typically, FEL pulses of higher energy per pulse exhibit a larger fraction of incoherent SASE radiation, which are unsuited for studies of coherent wave packet dynamics[14]. The result of the covariance analysis as a function of soft x-ray pump-probe delay is plotted in Fig. 3a.

Oscillations of the detected signal in the time domain are clearly visible and have been discussed in detail in ref.[8]. Supported by ab-initio simulations the observed short-time dynamics monitors a many-body quantum mechanical wave packet propagation, which is represented by a coherent superposition of electronic states dressed by vibrational excitations.

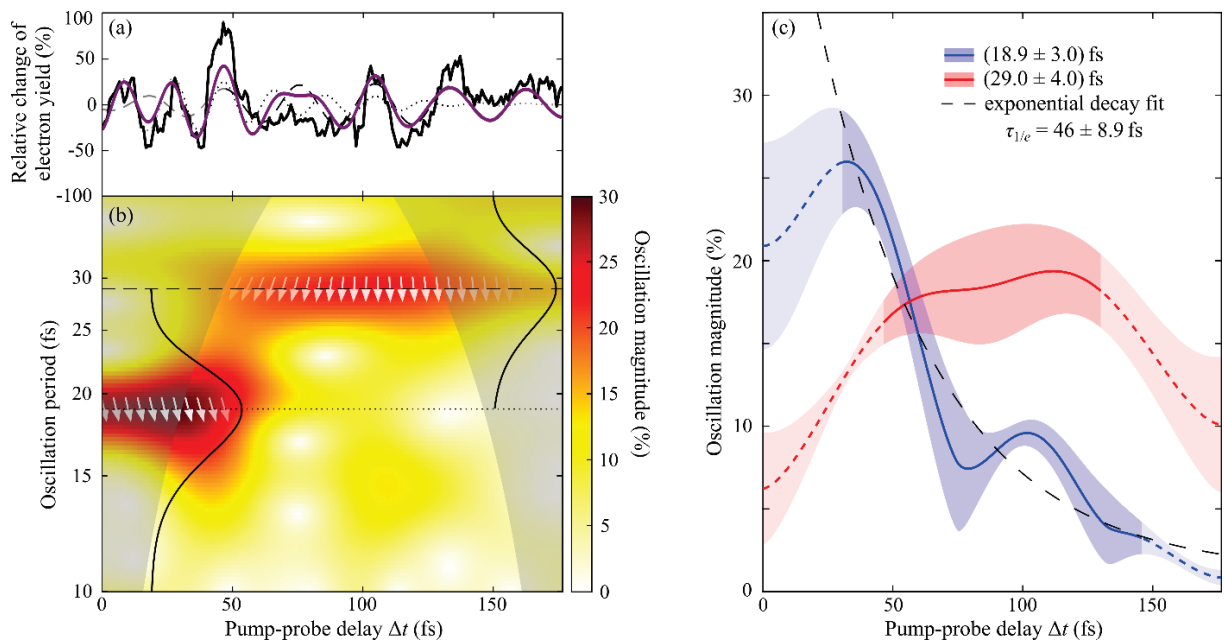

**Figure 3.** (a) Photoelectron - Auger electron - parent ion covariance signal as a function of pump-probe delay (black line). (b) Time-period distribution of correlated $Gly^{2+}$ ions with 247 eV < $E_{kin,1}$ < 257 eV and 10 eV < $E_{kin,2}$ < 210 eV electron multi-particle events derived from the wavelet analysis of the recorded signal trace. Two significant oscillations at (18.9 ± 3.0) fs and (29.0 ± 4.0) fs are observed (Gaussians) and marked as dashed and dotted lines. The nonlinear two-photon excitation scheme allows to retrieve the instantaneous phase of the many-body wave packet (white arrows). The oscillation amplitudes along the horizontal dashed and dotted lines marked in panel (b) are plotted in panel (c). The superposition of the two dominant oscillatory contributions making use of the measured amplitude, frequency and phase information is shown together with the recorded signal trace in panel (a) as violet line.

In the present work the recorded time-frequency spectrum of a quantum wave packet describing the electronic and nuclear motion is further analyzed by reconstructing the amplitude, the instantaneous phase and the instantaneous frequency of the modulus squared of the underlying total molecular wave function $|\Phi(\boldsymbol{r}, \boldsymbol{R}, t)|^2$ or more precisely of the modulus squared of the excited coherent superposition of eigenstates forming a many-body quantum wave packet in the intact glycine cation. In particular, we focus on the fate of the initially excited electronic coherence and the subsequent coupling to nuclear degrees of freedom. The Gly$^{2+}$ dication and covariant photo/Auger electron yield, which is plotted in Fig. 3a, shows a beating pattern and transition from (18.9 ± 3.0) fs to (29.0 ± 4.0) fs as evidenced by the time-period analysis by means of continuous wavelet transform (CWT). In general, a wavelet analysis allows to study the amplitude evolution of a non-stationary signal at scaling frequencies[15]. The oscillation frequencies were converted to periods for better comparability. The periods and standard deviation values were obtained by fitting a normal distribution to the peak in the first 50 fs and the peak between 75 fs to 175 fs respectively, as shown by the black solid line in Fig. 3b. Note that uncertainties of the analysis arise, where part of the wavelet (in time domain) extends past the finite recorded signal trace. The boundary for the start of these uncertainties is called 'cone of influence' (COI). Here, the COI is chosen as the points, where the autocorrelation magnitude of the respective wavelet decays by $1/e$ as indicated by the grey shaded area in Fig. 3b[16]. The extracted central periods are indicated in Fig. 3b as dashed and dotted horizontal lines throughout the full pump-probe delay range. It can also be seen, that the maximum at $T$ = 18.9 fs slightly shifts toward a 20-fs period after the first ~50 fs time delay, where its amplitude declines, which is attributed to the rising edge of the 29-fs oscillation. The experimental data shown in Fig. 3a can be reasonably well reconstructed by a simple superposition of two sine curves, each with a constant period and phase, but varying amplitude. The oscillation amplitudes derived from the wavelet analysis along the dashed and dotted lines in Fig. 3b are shown separately in Fig. 3c. The amplitude of the initial oscillation at early times with $T = (18.9 \pm 3.0)$ fs generally decays exponentially with a decay constant of $\tau_{1/e} = (46 \pm 8.9)$ fs indicating the timespan during which the electronic coherence in the glycine cation will be lost [17, 18]. Concurrently, the subsequent signal oscillation with time period $T = 29.0$ fs rises during the first 100 fs and declines afterward. It is attributed to a characteristic vibration of the nuclear configuration, due to the coupling of electronic to nuclear motion with time. The decreasing amplitude of the nuclear wave packet observed for long time delays reflects the final vibrational relaxation during energy dissipation.

The above picture indicates that here nuclear motion cannot be separated from the electronic one by applying the Born-Oppenheimer approximation. As we have discussed in detail in the introduction the total molecular wavefunction does not factorize for inner-valence photoionization with fs soft x-ray pulses, where the Born-Oppenheimer approximation breaks down. Describing the nonadiabatic processes at work in photoionized glycine compels us to revisit the basic representation of the time-dependent total molecular wave function (eq. 1), taking into consideration the complex interplay between nuclear and electronic motion[2]. As a starting point, we visualize the modulus squared of the excited coherent superposition of cationic eigenstates $|\Phi(r, R, t)|^2$ by using the amplitudes, the instantaneous phases and the instantaneous frequencies of the propagating many-body wave packet derived from the experimental data shown in Fig. 3. Note the experimental pump-probe scheme sketched in Fig. 1 with the probe-induced Auger decay as the key experimental observable follows a nonlinear two-photon process. This allows us to retrieve absolute phase information of the excited many-body quantum wave packet. The multi-dimensional character of $\Phi(r, R, t)$ makes it a challenge to visualize (see eq. 1), but we use in the present work a simple picture of a positive charge density, which moves as a function of a generalized electronic coordinate $r$ between the two carbon atoms (C, C$_\alpha$) and a generalized nuclear configuration coordinate $R$ describing the nuclear backbone geometry of the molecule between an inner and an outer turning point ($R_{min}$, $R_{max}$). One may want to argue that this one-dimensional visualization of the wave packet propagation in each coordinate according to the experimentally derived wave packet parameters (amplitudes, instantaneous frequencies and phases) is oversimplified to describe charge dynamics in polyatomic molecules, which is certainly true. Without detailed calculations, we cannot assign the configuration coordinate $R$ to a particular covalent bond elongation, such as the C$_\alpha$-C stretch. However, as we will show in the following this representation of the experimental data helps to understand the main features of the molecular wave function and its quantum dynamics observed in the soft x-ray absorption experiments at the carbon K-edge.

First of all, we have to recall that the recorded Auger yield, which enters the covariance plotted in Fig. 3a as a function of pump-probe delay, is a local measure of the time-dependent positive charge density in the vicinity of the carbon C$_\alpha$ atom[19]. This is because the cross-sections for transitions are proportional to the projection of the total molecular wave function onto the final state, i.e. the recorded Auger yield determines the transition probability from the local carbon 1s core shell (5a′ orbital) into the inner valence 10 a′ energy band formed by the corresponding 1h and 2h1p electronic configurations, as well as by further shake-up satellites. The pump-induced coherent superposition of these cationic eigenstates creates a many-body quantum wave packet, which propagates in space ($r$, $R$) and time $t$ in the intact glycine cation. Its amplitudes, instantaneous phases and frequencies are shown in Fig. 3. Thereby, it is possible to reconstruct and visualize the oscillatory charge motion and spreading of the electron hole depending on the electronic and nuclear coordinates described by the corresponding probability density $|\Phi(\boldsymbol{r}, \boldsymbol{R}, t)|^2$. Snapshots of the positive charge distribution at characteristic times along the recorded pump-probe trace, i.e. during the wave packet propagation, are given in Fig. 4. The amplitude ratio of the two dominant oscillations (18.9 ± 3.0) fs and (29.0 ± 4.0) fs is encoded in the color scale from blue at short pump-probe delays to red at long delays. The initial dynamics is plotted as vertical $r$- dependence, whereas the slightly longer time period is plotted as a function of $R$. The recorded signal strength of each oscillation is a measure of the transition dipole moment (5a′→10a′), since it determines the probe-induced Auger electron yield.

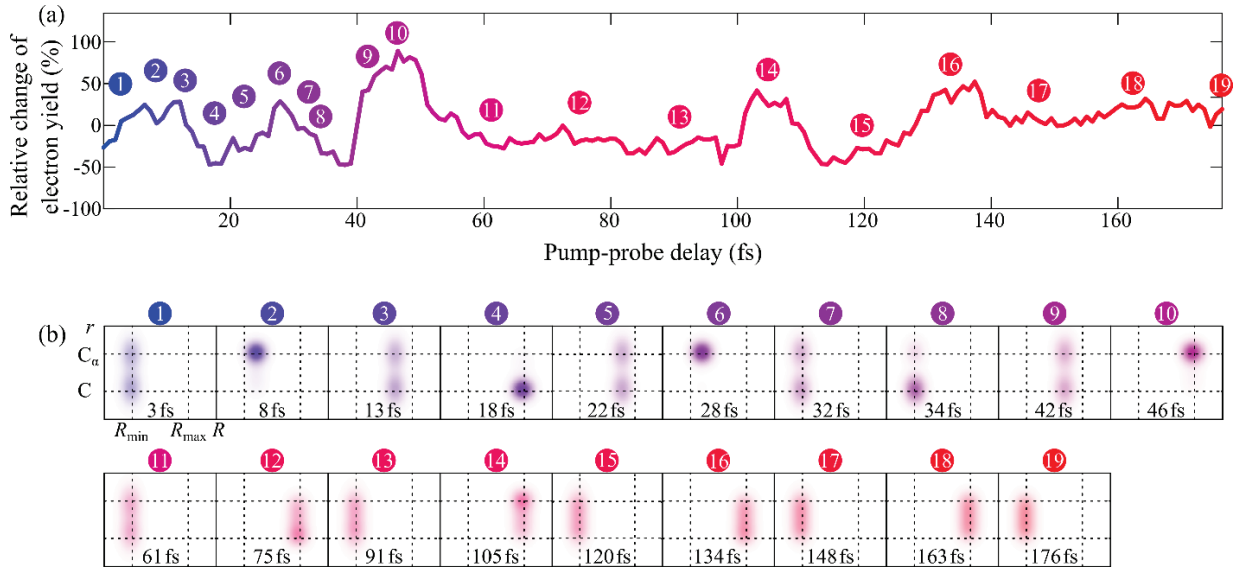

**Figure 4.** (a) Propagating many-body quantum wave packet during its lifecycle. (b) Snapshots of the positive charge distribution at characteristic times along the recorded pump-probe trace.

It is important to note that the underlying transition dipole moment has both an electronic and nuclear configuration coordinate origin. In the electronic coordinate the absorption is proportional to the local charge density at the $C_\alpha$ atom. One can clearly see that it takes several fs until maximum charge localization is reached at the $C_\alpha$ atom (motion picture 2 in Fig. 4), when starting from a coherent superposition of delocalized inner-valence electron eigenstates at $t = 0$. It is fascinating to see that the experimentally derived instantaneous phases of the many-body wave packet also prove that the quantum dynamics is initiated at the inner turning point $R_{min}$ of a characteristic configuration coordinate $R$. Obviously, soft x-ray photoionization out of a bonding orbital of the neutral Gly molecule results in a cation, which has its equilibrium geometry at larger intramolecular distances compared to the neutral ground state. At motion picture 4 the local positive charge density at the $C_\alpha$ atom is small and therefor the 5a′→10a′ transition probability and Auger signal strength is small although the nuclear configuration has reached its outer turning point $R_{max}$. An interesting situation in the dynamic interplay between electronic and nuclear motion is observed at motion picture 10 in the pump-probe data. Here, both the positive charge is localized at the $C_\alpha$ atom and the nuclear configuration is at the outer turning point. This scenario leads to a strong signal enhancement because the total transition dipole moment is maximized. In the further course of the wave packet propagation the electronic coherence is lost and the dynamics of the quantum system is exclusively governed by nuclear motion in the $R$ coordinate until vibrational relaxation dissipate the excitation energy among the 3$M$-6 nuclear degrees of freedom or molecular fragmentation sets in ref.[20].

CONCLUSIONS

Nonadiabatic quantum wave packet dynamics during charge propagation in the amino acid glycine is visualized by reconstructing the oscillatory charge density distribution as a function of a characteristic electron coordinate *r* and a time-dependent nuclear configuration coordinate *R* from time-resolved Auger electron spectroscopic data using fs x-ray pulses. Snapshots of the dynamic interplay between strongly coupled electronic and nuclear motion described by the modulus squared of the excited coherent superposition of eigenstates are given at characteristic times along the recorded pump-probe trace, i.e. during the excited many-body quantum wave packet propagation. The analysis of its amplitudes, instantaneous phases and instantaneous frequencies provides unprecedented insight into quantum dynamical processes beyond the Born-Oppenheimer approximation. The current two-dimensional representation of the experimental data is a simplification, that has the possibility to be generalized to other cases, but further full calculations of the coupled electron-nuclear dynamics in glycine are needed to gain additional insight. Recently, interesting results have been reported by Lara-Astiaso et al.[21] In their work, they focused on the impact of electron-nuclear coupled dynamics on charge migration in glycine induced by attosecond pulses. The time-dependent density functional theory (TDDFT)-Ehrenfest calculations consider electron correlation and the interplay between electron and nuclear dynamics during time propagation. However, a direct comparison with our experimental data is somewhat difficult, since in the present experiments, small-bandwidth, soft x-ray FEL pulses at 273 eV have been applied, whereas the theoretical work focused on broadband XUV photon pulse excitation typical for high-harmonic generation (HHG) sources. These pulses have a bandwidth of approximately 19 eV, spanning from 16 to 35 eV photon energies. As a result, they can ionize the molecule from the highest occupied molecular orbital (HOMO) down to the HOMO-14, encompassing fifteen valence and inner-valence molecular orbitals, which result in much broader wave packets and, thus, shorter time periods of the induced coherences.

AUTHOR INFORMATION

**Corresponding Author**

*tim.laarmann@desy.de.

**Author Contributions**

The manuscript was written through contributions of all authors. All authors have given approval to the final version of the manuscript.

**Notes**

The authors declare no competing financial interest.

ACKNOWLEDGMENT

We acknowledge DESY (Hamburg, Germany), a member of the Helmholtz Association HGF, for the provision of experimental facilities. The experimental part of this research was carried out at FLASH, beamline FL24. Beamtime was allocated for Proposal No. F-20191551. This work was funded by the Deutsche Forschungsgemeinschaft (DFG, German Research Foundation) through the Cluster of Excellence "Advanced Imaging of Matter" (EXC 2056, project ID 390715994) and Project Nos. KI 482/20-2 and LA 1431/5-2. J.P.M. acknowledges financial support from the UK Science and Technology Facilities Council and the UK Engineering and Physical Sciences Research Council through Grant Nos. EP/X026094/1, EP/T006943/1 and EP/V026690/1. We also acknowledge the scientific exchange and support of the Center for Molecular Water Science (CMWS). We acknowledge the use of the Maxwell computational resources operated at Deutsches Elektronen-Synchrotron DESY, Hamburg, Germany.